\documentclass[aps,jap,twocolumn,superscriptaddress,a4paper]{revtex4}

\usepackage{graphicx}
\usepackage[colorlinks=true,citecolor=blue,linkcolor=magenta]{hyperref}
\usepackage{amsmath}
\hypersetup{
pdfauthor = {Patrick Maletinsky},
colorlinks = true, linkcolor = blue, urlcolor=blue, bookmarksnumbered =  true}

\newcommand{\ket}[1]{\left|#1\right>}

\begin{document}
%Title of paper
\title{Fabrication of all diamond scanning probes for nanoscale magnetometry }
\affiliation{Department of Physics, University of Basel, Klingelbergstrasse 82, Basel CH-4056, Switzerland}
\affiliation{Experimentalphysik, Universit\"at des Saarlandes, D-66123 Saarbr\"ucken, Germany}
\affiliation{These authors contributed equally}

\author{Patrick Appel}
\affiliation{Department of Physics, University of Basel, Klingelbergstrasse 82, Basel CH-4056, Switzerland}
\affiliation{These authors contributed equally}
\author{Elke Neu}
\affiliation{Department of Physics, University of Basel, Klingelbergstrasse 82, Basel CH-4056, Switzerland}
\affiliation{Experimentalphysik, Universit\"at des Saarlandes, D-66123 Saarbr\"ucken, Germany}
\affiliation{These authors contributed equally}
\author{Marc Ganzhorn}
\affiliation{Department of Physics, University of Basel, Klingelbergstrasse 82, Basel CH-4056, Switzerland}
\affiliation{These authors contributed equally}
\author{Arne Barfuss}
\affiliation{Department of Physics, University of Basel, Klingelbergstrasse 82, Basel CH-4056, Switzerland}
\author{Marietta Batzer}
\affiliation{Department of Physics, University of Basel, Klingelbergstrasse 82, Basel CH-4056, Switzerland}
\author{Micha Gratz}
\affiliation{Experimentalphysik, Universit\"at des Saarlandes, D-66123 Saarbr\"ucken, Germany}
\author{Andreas Tsch\"ope}
\affiliation{Experimentalphysik, Universit\"at des Saarlandes, D-66123 Saarbr\"ucken, Germany}
\author{Patrick Maletinsky}
\email[]{patrick.maletinsky@unibas.ch}
\affiliation{Department of Physics, University of Basel, Klingelbergstrasse 82, Basel CH-4056, Switzerland}

%\homepage[]{Your web page}
%\thanks{}

\date{\today}

\begin{abstract}
 The electronic spin of the nitrogen vacancy (NV) center in diamond forms an atomically sized, highly sensitive sensor for magnetic fields. To harness the full potential of individual NV centers for sensing with high sensitivity and nanoscale spatial resolution, NV centers have to be incorporated into scanning probe structures enabling controlled scanning in close proximity to the sample surface. Here, we present an optimized procedure to fabricate single-crystal, all-diamond scanning probes starting from commercially available diamond and show a highly efficient and robust approach for integrating these devices in a generic atomic force microscope. Our scanning probes consisting of a scanning nanopillar (200 nm diameter, $1-2\,\mu$m length) on a thin ($< 1\mu$m) cantilever structure, enable efficient light extraction from diamond in combination with a high magnetic field sensitivity ($\mathrm{\eta_{AC}}\approx50\pm20\,\mathrm{nT}/\sqrt{\mathrm{Hz}}$). As a first application of our scanning probes, we image the magnetic stray field of a single Ni nanorod. We show that this stray field can be approximated by a single dipole and estimate the NV-to-sample distance to a few tens of nanometer, which sets the achievable resolution of our scanning probes.
\end{abstract}

\maketitle

\section{Introduction \label{sec:Int}}

%Motivation and background
The negatively charged nitrogen vacancy center (NV center) in diamond forms a highly promising sensor: On the one hand, its unique combination of long spin coherence times and efficient optical spin readout enables the detection of magnetic \cite{Maze2008} and electric fields \cite{Dolde2011} as well as local temperature.\cite{Toyli2013a,Acosta2010b} On the other hand, the NV center is a highly photostable single photon source and therefore an ideal emitter for scanning near field \cite{Tisler2013a} and single photon microscopy.\cite{Sekatskii1996} Moreover, all properties relevant for sensing are sustained from cryogenic temperatures \cite{Thiel2015, Pelliccione2014} up to $550\,$K,\cite{Toyli2012} rendering NV centers highly promising not only for applications in material sciences and physics but also for applications in the life sciences.\cite{LeSage2013} As a point defect in the diamond lattice, the NV center can be considered as an 'artificial atom' with sub-nanometer size. As such, it promises not only highest sensitivity and versatility but in principle also unprecedented nanoscale spatial resolution.\

Triggered by this multitude of possible applications, various approaches to bring a scanable NV center in close proximity to a sample were recently developed. The first experiments in scanning NV magnetometry employed nanodiamonds (NDs) grafted to atomic force microscope (AFM) tips.\cite{Balasubramanian2008,Rondin2012,Rondin2014,Tetienne2014} However, NVs in NDs suffer from short coherence times limiting their sensitivity as a magnetic sensor. Secondly, efficient light collection from NDs on scanning probe tips is difficult and limits the resulting sensitivities. Lastly, it has proven challenging to ensure close NV-to-sample separations in this approach. Most published work reported on NDs scanning within $\gtrsim100~$nm from the sample surface, limiting the spatial resolution of the scanning probe imaging. Additionally, the emission of NV centers in single digit NDs is typically unstable without further treatment.\cite{Bradac2010} Motivated by these drawbacks, a novel approach using all-diamond, single crystalline AFM tips has recently been demonstrated.\cite{Maletinsky2012} This approach relies on fabricating scanning probes with the NV center placed close to the apex of a scanning diamond nanopillar. Beside close proximity of the NV center to the sample, the pillar's light guiding properties enhance collection efficiency for the NV fluorescence and the devices can be sculpted out of high purity diamond, which enables long coherence times. Thus, color centers with optimal properties (regarding photo-stability and spin-coherence) in high purity material and efficient light collection can be used as sensors.

%Summary
In this paper, we describe an optimized procedure to fabricate such single-crystal, all-diamond scanning probes. In particular, we present in detail the nanofabrication of diamond nanopillars for scanning probe microscopy and describe a highly efficient and robust approach for integrating these devices in an atomic force microscope (AFM). We discuss the magnetometry performance of the probes and demonstrate high resolution imaging of the stray field of single magnetic Ni nanorods using the all-diamond scanning probes.

\section{Fabrication of all diamond scanning probes}

The fabrication procedure that we describe here consists of 6 steps: We start with commercially available, high purity diamond plates ($50~\mathrm{\mu}$m thick, Section \ref{sec:initialstep}) in which we create shallow NV centers using ion implantation (Section \ref{sec:NVcreation}). Our all diamond scanning probes consist of a cylindrical nanopillar ($200~$nm diameter, $1.5~\mathrm{\mu}$m height) on a $<1~\mathrm{\mu}$m thick cantilever. Thus, it is essential to thin down the commercially available plates to a suitable thickness (Section \ref{sec:deepetch}). The thinned membranes are subjected to two consecutive lithography and plasma etching steps to form the pillars and the cantilever (Section \ref{struct_scan}). In the subsequent step, we identify the scanning probes that contain single NV centers (Section \ref{sec:precharacterization}). Finally, we mount the selected scanning probes to a tuning fork based AFM head (Section \ref{sec:transfer}).

\subsection{Diamond material and initial sample preparation \label{sec:initialstep}}
Our nano-fabrication procedure for the all-diamond scanning probe devices is based on commercially available,  high purity, synthetic diamond grown by chemical vapor deposition (Element Six, electronic grade, [N]${}^s$$<$$5~$ppb, B$<$$1~$ppb).\cite{e6url} The $500~\mu$m thick diamonds are processed into 30-$100~\mu$m thick diamond plates by laser cutting and subsequent polishing (Delaware Diamond Knives, USA or Almax Easy Lab, Belgium \cite{Almaxurl}). While our process can be applied to a large range of thicknesses, we found $50~\mu$m thick plates to form the best compromise between mechanical stability, ease of handling and reasonable processing times (see Section \ref{sec:deepetch}).

The surface roughness of the starting diamond plates is typically $0.7~$nm, as evidenced by AFM imaging [Fig.\ \ref{Fig:membranes}(d)], and the plates have a wedge of typically several micrometers across the lateral sample dimensions of $4~$mm. We note that such a high quality polish is mandatory for the subsequent processing steps. Initially, we clean the plates using a boiling tri-acid mixture (1:1:1 sulfuric acid, perchloric acid, nitric acid, boil acid mixture until reverts to clear appearance) to remove any surface contamination which might have resulted from polishing.\cite{Hird2004,Schuelke2013} Lastly, the sample is cleaned in solvents (deionized water, acetone, ethanol, isopropanol) to remove possible contaminants present in the acids.

Mechanical polishing of diamond is known to introduce crystal damage below the polished surface into a depth of up to several micrometers.\cite{Volpe2009, Friel2009, Naamoun2012} The lattice in this highly damaged layer can be strongly deformed and defective: cathodoluminescence (CL) measurements indicate a high concentration of defects\cite{Volpe2009} and etching away 3-$4~\mu$m of diamond almost recovers the CL of pristine diamond.  NVs in this damaged layer might therefore suffer from a unstable charge state or spin decoherence due to trapped paramagnetic defects or fluctuating charges. Furthermore, the highly strained layer might render the NV spins insensitive to magnetic fields in first order and therefore useless for magnetometry.\cite{Rondin2014} To circumvent these potential obstacles, we remove $\approx 3~\mu$m or more of the damaged surface layer using inductively coupled reactive ion etching (ICP-RIE) as described in the following.

For all etch steps, the diamond plates are mounted on Si chips ($1~$cm squares) as carriers; we perform plasma etching using a Sentech SI 500 ICP-RIE apparatus. We initiate the etching by removing roughly the first micrometer of diamond using an ArCl$_2$ plasma step. This plasma chemistry has been reported to remove damaged diamond layers without roughening the surface.\cite{Friel2009} Note that even slight surface roughening would be detrimental for all subsequent processes. We summarize the plasma parameters used as well as the resulting etch rates [as determined by an in-situ laser interferometer (SenTech SLI 670)] in table \ref{tab:plasma}. While enabling optimal etching of defective diamond, the ArCl$_2$ plasma also strongly erodes Si carrier wafers routinely used in ICP-RIE processes. The resulting high level of Si contamination introduces a roughening of the diamond surface. To avoid this, we employ a ceramics based carrier system which we find to be more resistant to etching in the ArCl$_2$ plasma consequently avoiding contamination. Diamond surfaces prepared by ArCl$_2$ plasma have been suspected to contain Cl$_2$,\cite{Tao2014} which might deteriorate the NV spin properties. As a consequence, we terminate etching using an O$_2$ plasma to remove any such potential Cl$_2$ contamination (see  table \ref{tab:plasma}).
\begin{table}
\caption{\label{tab:plasma}Plasma parameters for the nano-fabrication procedure. Note that the ArO$_2$ plasma is used to etch the nanopillar structures, while the other plasma types are used for the 'deep etches' to remove polishing damage and form the thin membrane. The nanopillar etching is carried out using a ($6~$Inch) silicon carrier inside the reactor, while all other etches are performed using a ceramics carrier (96\% Al$_2$O$_3$) to avoid silicon contamination. The plasma bias voltage was stable within roughly 10\% for runs performed within a time-span of several weeks.}
\begin{ruledtabular}
\begin{tabular}{cccp{1cm}cc}
plasma  & ICP power & RF power/bias & flux & Pressure & Etch rate \\
  & [W] & [W]/[V] & [sccm] & [Pa] & [nm/min] \\
\hline
ArCl$_2$ & 400  & 100/220 & Ar 25 \newline Cl$_2$ 40  & 1  & 60  \\
O$_2$ & 700  & 50/120 & O$_2$ 60  & 1.3  & 150  \\
ArO$_2$ & 500  & 200/120 & Ar 50 \newline O$_2$ 50  & 0.5  & 150  \\
\end{tabular}
\end{ruledtabular}
\end{table}
\subsection{Creation of NV color centers \label{sec:NVcreation}}
To realize high resolution imaging, it is mandatory to achieve close proximity between NV spin and sample, which implies the creation of NV centers close to the diamond surface. To create such a shallow layer of NV centers, we implant the etched diamond surface with ${}^{14}$N ions at an energy of 6 keV and a dose of $3\times 10^{11}~$cm$^{-2}$ (Ion beam services, France). The estimated resulting stopping depth of the  ${}^{14}$N ions in diamond is $9\pm4$ nm.\cite{SRIM} We anneal the sample in vacuum (chamber base pressure: 3-$4\times10^{-7}$ mbar) partly following the recipe from Ref.\ \onlinecite{Chu2014}. The heating device is a boron nitride plate, directly, electrically heated via buried graphite strips (Tectra, Boralectric HTR-1001). The temperature of the oven is calibrated using a comparison between pyrometer measurements and a thermocouple (tungsten/rhenium) inserted into a bore hole in the heater plate. We use the following sequence of annealing steps: ramp in $1~$h from room temperature to 400$^\circ$C, hold $4~$h at 400$^\circ$C, ramp in $1~$h to 800$^\circ$C, hold at 800$^\circ$C for $2~$h, cool down. We also investigated the effect of a high temperature annealing step at 1200$^\circ$C (ramp in $1~$h 800$^\circ$C to 1200$^\circ$C, hold at 1200$^\circ$C for $2~$h) according to Ref.\ \onlinecite{Chu2014}. However, we did not find any significant effect on the NV yield or the NV spin coherence properties.
With the previously described procedure, we create a layer of NV centers with a density of $2.6\times 10^9\, \mathrm{cm}^{-2}$ (see Section \ref{sec:precharacterization}). From this, we estimate the yield of the NV creation to be $0.9\,\%$ which is comparable to previously reported values.\cite{Pezzagna2010}

\subsection{Deep Etching to form diamond membranes \label{sec:deepetch}}
\begin{figure}
\includegraphics[width =8.6cm]{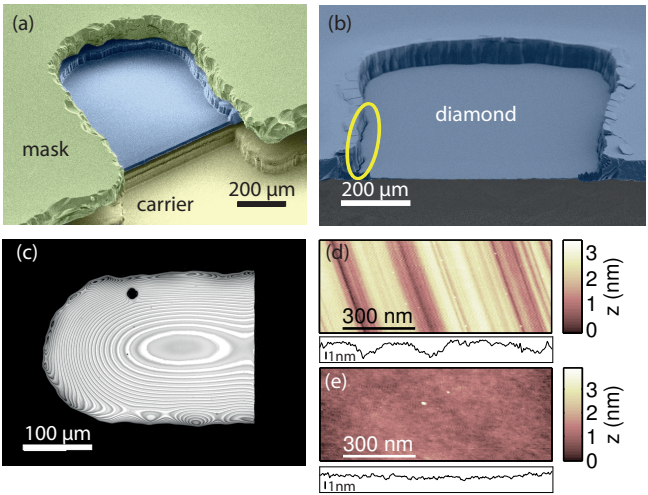}
\caption{(a) False-colored scanning electron microscopy (SEM) image of a diamond plate (blue) sandwiched between a quartz cover slip (green) and a Si carrier (yellow). (b) Thin diamond membrane etched in one run (c) Laser scanning confocal microscope (Keyence VK-X210, $\lambda=408~$nm) image of a 10-$12~\mu$m thick membrane etched from a $57~\mu$m thick plate. Note the interference fringes, witnessing a thickness variation of roughly $1~\mu$m (middle to sidewall) as well as the formation of a trench close to the mask. (d) AFM measurement of the commercially available diamond plate; marks due to polishing are clearly visible, the RMS roughness amounts to $0.7~$nm. (e) AFM image of the same plate thinned to form a micron-thick membrane, the roughness is reduced ($0.3~$nm) and the polishing marks are no longer visible.   \label{Fig:membranes}}
\end{figure}
We now introduce an etching process leading to a thinned membrane of several micron thickness and of around $400\times400~\mu$m size supported by the surrounding $50~\mu$m thick diamond plate. Typical etch masks with sub-micron thickness would not withstand the long etching process necessary to thin a $50~\mu$m thick diamond plate down to a few microns. Thus, we employ thin quartz cover slips (SPI supplies, 75-$125~\mu$m thick) as etch masks. Using water jet cutting (Microwater Jet, Switzerland) a slot ($\leq 500~\mu$m width) is cut into the cover slip. The sample is then sandwiched between a Si carrier chip and the mask; the latter is fixed onto the 6 inch carrier wafer using vacuum grease [see Fig.\ \ref{Fig:membranes}(a)]. The etch resistance of the quartz material allows for a high quality etching, whereas using standard glass cover slips leads to micro-masking and roughening of the etched diamond as a results of low etch resistance. The masks can be reused several times.

For the membrane 'deep etch', we use an ArCl$_2$ and an O$_2$ based plasma, with plasma parameters as summarized in table \ref{tab:plasma}. The etching process starts with $5~$mins of ArCl$_2$ plasma, then the following sequence is cycled until the desired etch depth is reached: 5 mins ArCl$_2$, $5~$mins O$_2$, $5~$mins O$_2$. Consecutive etch steps were separated by 5 mins of cooling under Ar ($100~$sccm, $13.2~$Pa). In the ICP-RIE plasma, a trench forms close to the edge of the quartz mask and the sidewalls of the pit etched into the diamond plate, see yellow marker in \ref{Fig:nanofab}(b). As the depth of this trench can exceed $1~\mu$m during our deep etch, the thinned membrane becomes mechanically unstable as its connection to the thick diamond plate is compromised. The formation of the trench can be explained as follows: the reflection of high energy ions impinging under grazing incidence onto the sidewalls of the mask and the already etched pit leads to a focusing of the ions close to the sidewalls of the pit and a locally enhanced etch rate induces the trench.\cite{Hoekstra1998}  To ensure membrane stability, we exchange the initial etch mask (mostly 400-$500~\mu$m etched area) for a narrower mask (300-$400~\mu$m) when the membrane has reached a thickness of about 8-$10~\mu$m. Due to the shifted mask edge, the trench formation restarts at the new mask edge location [see e.g.\ Fig.\ \ref{Fig:nanofab}(b), right side]. The trench formed during the residual etching does not destabilize the membrane.

Due to the thick etch mask, we observe a significantly non uniform thickness of the final membrane, which is much thicker close to the mask than in the center. We measure the membrane's thickness at its free-standing edge using an SEM and estimate the overall thickness variation using a laser scanning confocal microscope [see Fig.\ \ref{Fig:membranes}(c)]. Our membranes for scanning probe fabrication finally have a thickness of around $1.5~\mu$m in the center and 2.5-$3~\mu$m close to the mask. AFM measurements show that the etching process improves the surface quality of the membrane: Polishing marks observed before the etching [Fig.\ \ref{Fig:membranes}(d), RMS roughness $0.7~$nm] are not observed anymore after the deep etch  [see Fig.\ \ref{Fig:membranes}(e)] and we find an RMS roughness of $0.3~$nm for the thinned membrane.

We note that the trenching at the rim of the membranes as well as the non-uniformity might be reduced or even avoided using quartz masks with angled sidewalls. Such angled sidewalls could reduce the effective thickness of the mask and thus lead to a more uniform etch rate and less trenching. Deep etches using this novel mask geometry engineered using laser cutting (Photonikzentrum Kaiserslautern, Germany) are currently being investigated.

\subsection{Structuring Scanning Probes \label{struct_scan}}
 \begin{figure}
\includegraphics[width =8.6cm]{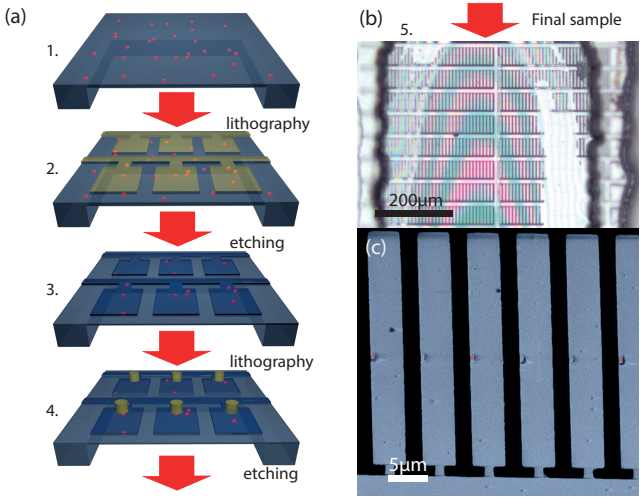}
\caption{(a) Schematic of the fabrication process for the all diamond scanning probes; starting from the membrane (1.) with the shallow implanted NV centers (red). We use electron beam lithography to structure a pattern (2.) consisting of transverse holding bars to which $20~\mu$m long,  $3~\mu$m wide cantilevers are connected via $500~$ nm bridges [see also see Fig.\ \ref{Fig:characterization}(a)]. The structure is transferred in diamond (3.) via ICP RIE and afterward we sculpt the pillar on top via lithography (4.) and subsequent etching (5.); (b) Optical and (c) SEM image of fabricated pattern of all diamond scanning probes fabricated using the alternative approach described at the end of section \ref{struct_scan}. \label{Fig:nanofab}}
\end{figure}

  Our scanning probes consist of a $20~\mu$m long,  $3~\mu$m wide cantilever, which holds a nanopillar for scanning and sensing [see Fig.\ \ref{Fig:nanofab}(c) and Fig.\ \ref{Fig:characterization}(a)]. Following Ref.\ \onlinecite{Babinec2010}, we aim for pillars with $\approx200~$nm diameter and a straight, cylindrical shape to enable efficient collection of the NV fluorescence.  The cantilevers are connected to a holding bar in the membrane by $500~$nm wide bridges. These bridges are strong enough to reliably fix the cantilever to the membrane, but still allow for easy breaking off of the cantilever for subsequent mounting onto an AFM head.\

  To form these scanning probes, we use two mutually aligned electron beam lithography steps each followed by structuring via ICP-RIE. In the first step, the holding bar pattern together with the cantilevers are formed. Subsequently, pillars are structured on top of the cantilevers, as sketched in Fig. \ref{Fig:nanofab}(a)\

For lithography, we use hydrogen silsesquioxane (HSQ) negative electron beam resist (FOX-16, Dow Corning) as an etch mask. To create a thick mask with a high aspect ratio, we evaporate $2~$nm  Ti as an adhesion layer before spin coating a $600~$nm thick layer of HSQ, which we bake on a hotplate at $90^{\circ}$C for $10~$min. Note that the Ti layer only efficiently enhances the adhesion when not allowed to oxidize before applying the resist. We use electron beam lithography with $30~$keV to pattern the HSQ layer. To prevent charging of the diamond sample, we expose the mask with currents below $50~$pA and structure our $200~$nm diameter pillar with a dose of $1500~\mu$As/cm$^2$ and the cantilever with a dose of $150~\mu$As/cm$^2$. Finally, we develop the samples for $20~$s in $25~$wt\% TMAH and remove the Ti in $70^{\circ}$C hot 37\% HCl. Both steps are followed by rinsing in de-ionized water and cleaning in isopropanol.\

We transfer the HSQ masks into the diamond via an ArO$_2$ plasma (parameters see table \ref{tab:plasma}). Our ArO$_2$ plasma enables a highly anisotropic etch while simultaneously creating a smooth surface in-between the etch masks. After each etch step, we remove residual HSQ and Ti using 20:1 buffered oxide etch (10:10:1 deionized water, ammonium fluoride,  40\% HF) and clean the sample in a boiling tri-acid mixture and a solvent clean (see Section \ref{sec:initialstep}).

Fabricating the scanning probes requires multiple steps as illustrated in  Fig.\ \ref{Fig:nanofab}(a): In the first step, we structure the pattern consisting of the transverse holding bars and the cantilevers. Additionally, markers (crosses) located adjacent to the thin membrane are defined in the HSQ mask and transferred into the surrounding diamond plate simultaneously to the pattern [markers not shown in Fig.\ \ref{Fig:nanofab}(a)]. In the second step, we spin coat HSQ on top of the etched pattern which on top of the structures forms a homogeneous film. To ease marker identification, we mechanically remove the HSQ film on top of the markers. This allows us to clearly identify the markers during electron beam lithography and use them to align the pillars with respect to the cantilevers. In the last step, we transfer the pillar pattern into the diamond. As only the pillar is protected by an HSQ mask, the previously defined pattern including the membrane is thinned down during this etching.  We continue etching, until the membrane is thinned to a point where all diamond material in-between the cantilevers has been etched away and the cantilevers remain free-standing. Note that the length of the pillars is limited by mask erosion and faceting, as well as the formation of a trench around the pillar (see also Section \ref{sec:deepetch}) leading to detachment of the pillar from the cantilever. In general, we are able to etch $2\,\mu$m long wires with a $600\,$nm thick HSQ mask. As a consequence, we start with a membrane of $2-3\,\mu$m and etch $\sim 1\,\mu$m deep when we transfer the holding bars and cantilevers into the membrane. In the second step, we are thus able to etch $\sim 2\,\mu$m long pillars while removing all diamond material in-between the cantilevers. It should also be noted, that we have observed micromasking effects forming needles at the edge of the cantilever during this final etch step. While the magnetometry performance remains unaffected, we have explored an alternative approach to eliminate such micromasking effects: based on the work of Ref. \onlinecite{Maletinsky2012}, we have also structured the cantilevers and pillars from different sides of the membrane [examples shown in Fig. \ref{Fig:nanofab}(b) and (c)]. Although this approach fully eliminates the above mentioned micromasking problem, the alignment of the pillar with respect to the cantilever becomes challenging. Despite these drawbacks, both techniques allow to produce hundreds of scanning probes on a single membrane [see Fig. \ref{Fig:nanofab}(b)]

 Furthermore the nano-fabrication results we present have been obtained using (100) oriented diamond material, however first results clearly indicate that our fabrication process is not restricted to this crystal orientation and can be extended to orientations more favorable for NV sensing applications e.g.\ (111).\cite{Neu2014}\

\subsection{Device characterization \label{sec:precharacterization}}

\begin{figure}
\includegraphics[width =8.6cm]{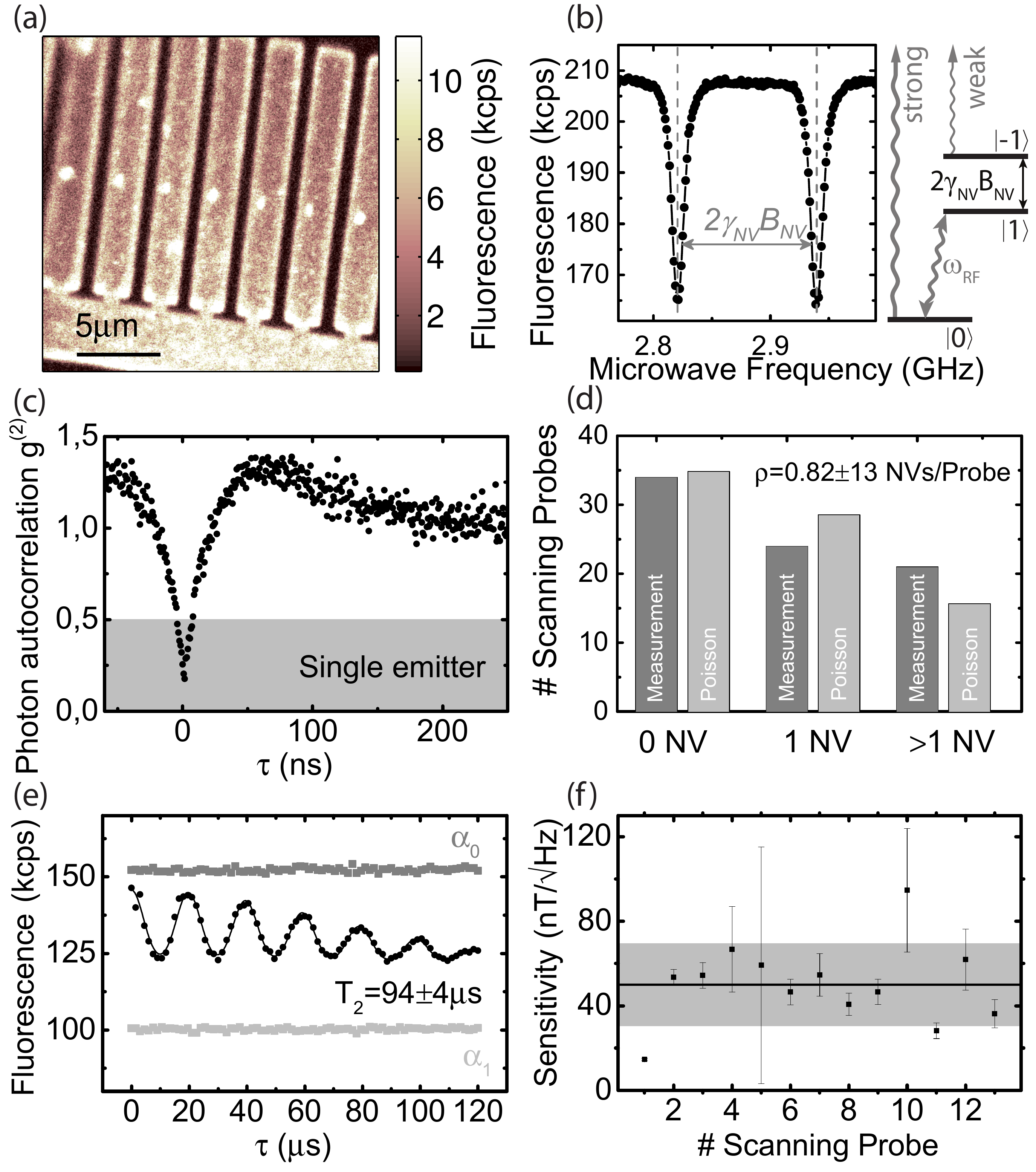}
\caption{(a) Confocal image of an array of all-diamond scanning probes in $10^3$ counts per second (kcps). (b) Typical optically detected electron spin resonance and schematic illustrating the electronic ground state spin configuration; the ground state spin triplet can be optically read out since the $ \ket{0} $ level posseses a higher fluorescence rate$\,\alpha_0$ compared to $ \ket{\pm1}$ ($\alpha_{\pm1}$). (c) Photon autocorrelation curve, $\mathrm{g^{(2)}}$, measured on a single NV center within a pillar. (d) Statistics of the NV number per scanning probes for 79 scanning probes together with a Poissonian fit yielding an average of $0.82\pm 0.13\,$ NV centers/scanning probe.(e) Hahn echo measurement from a single NV center in a scanning probe. The envelope fitted to the characteristic echo decay (see main text) yields a decay time $\mathrm{T_{2}}=94\pm 4\,\mu$s. (f) AC sensitivity for the 13 scanning probes with single NV centers, as determined from coherence times and optical readout contrasts (see main text). The black lines indicates the mean value of $\eta_{\rm AC}\approx 50\pm 20\,\mathrm{nT/\sqrt{Hz}}$ and the gray region illustrates the standard deviation of the sensitivities. . \label{Fig:characterization}}
\end{figure}

We characterize the scanning probes to identify the most suitable devices to be transferred and integrated into our AFM setup. For this, we employ a homebuilt confocal microscope equipped with microwave control electronics to perform electron spin resonance (ESR) and Hahn echo measurements to determine the NV spin coherence time $T_{2}$. Additionally, the setup is equipped with correlation electronics to perform second order autocorrelation ($g^{(2)}$) measurements to identify single NV centers.\

Figure\,\ref{Fig:characterization}(a) shows a confocal fluorescence map of our structured scanning probe array obtained by redording the photoluminescence (PL) in a spectral window above $550\,$nm. To identify the scanning probes with single and multiple NV centers, we measure the ESR spectra and $g^{(2)}$. Using a resonant microwave driving field, the NV center can be promoted from the $\ket{0}$ state to the less fluorescent $\ket{\pm 1}$ state, which allows for an efficient optical detection of NV ESR, as depicted for a single NV center in Fig. \ref{Fig:characterization}(b). A static magnetic field leads to a splitting $2 \gamma_{\rm NV} B_{\rm NV}$ of the two NV ESR resonances ($\ket{0}$ to $\ket{1}$ and $\ket{0}$ to $\ket{-1}$), where $\gamma_{\rm NV}=2.8\,\rm MHz/G$ is the gyromagnetic ratio and $B_{\rm NV}$ the magnetic field along the NV symmetry axis. Thus, scanning probes with multiple NV centers aligned along more than one of the four equivalent $<111>$ crystal-directions show multiple resonances. While multiple pairs of ESR dips quickly identify multiple NVs, no ESR signal identifies pillars without NV$^-$.  Scanning probes with single NV centers are reliably identified by a significant antibunching dip below $0.5$ in the $g^{(2)}$ measurement [see Fig.\,\ref{Fig:characterization}(c)]. Using these measurements, we classify the scanning probes into devices with no, single and multiple NV.\

Figure \ref{Fig:characterization}(d) shows the statistics of the number of NVs found in 79 scanning probes and reveals that approx. $30\,\%$ of them yield single NV centers. As expected, the number of NV centers per scanning probe follows a Poisson distribution. Using the probability for 0 and 1 NV center per pillar, we deduce an average number of NV centers of $0.82\pm 0.13\,$NV centers/scanning probe [see Fig.\ \ref{Fig:characterization} (d)] corresponding to a NV density of $2.6\times 10^9\, \mathrm{cm}^{-2}$ and a creation yield of $0.9\,\%$. We note that we observed a high variation of this value between different samples, which we attribute to variations of pillar diameters, uncertainty in the implanted nitrogen dose and possible variations in material properties (e.g. strain or vacancy concentrations).

 The magnetometry performance of scanning probes with single NV centers is typically characterized by their sensitivity $\eta$ to magnetic fields. The sensitivity set by the spin coherence properties of the NV center and the detected fluorescence rate in the $ \ket{0} $ and $ \ket{1} $ state can be derived from a Hahn-Echo measurement as depicted in Fig.\,\ref{Fig:characterization}(e). The data are fitted using the formula \cite{Childress2006a}

\begin{eqnarray}
F(\tau) &=& \frac{\alpha_\mathrm{0}+\alpha_\mathrm{1}}{2} \\\nonumber
&+&\frac{\alpha_\mathrm{0}-\alpha_\mathrm{1}}{2} \mathrm{exp}[-\left(\frac{\tau}{T_\mathrm{2}}\right)^n\sum_{j} \mathrm{exp}[-\left(\frac{\tau-j\tau_\mathrm{rev}}{T_\mathrm{dec}}\right)^2],
\end{eqnarray}

where $\alpha_\mathrm{0}$ and $\alpha_\mathrm{1}$ are the detected fluorescence rates of the NV in the $ \ket{0} $ and $ \ket{1} $ state respectively [see Fig.\,\ref{Fig:characterization}(b)] and $T_\mathrm{2}$ is the spin coherence time. The exponent $n$ depends on details of the decoherence process,\cite{Medford2012} whereas $\mathrm{\tau_{rev}}$ indicates the revival period associated with the Larmor precession of the $^{\rm 13}$C nuclear spins and $T_\mathrm{dec}$ the correlation time of the $^{\rm 13}$C nuclear spin bath.\cite{Childress2006a} For the depicted Hahn echo measurement, we derive $\alpha_1=98\pm1\,\mathrm{kcps}$, $\alpha_0=146\pm1\,\mathrm{kcps}$, $T_{2}=94\pm 4\,\mathrm{\mu s}$, $n=2.1 \pm 0.2$, $\tau_{\mathrm{rev}}=19.8 \pm 0.1\,\mathrm{ns}$ and $T_{\mathrm{dec}}=5.9 \pm 0.2\,\mathrm{ns}$. Note that the detected fluorescence rates are a factor of $\sim 3$ higher compared to shallow implanted NV centers in unstructured samples due to fluorescence waveguiding in the pillar.

%The magnetic field sensitivity $\mathrm{\eta_{AC}}$ of a given scanning probe with a single NV center depends on the signal to noise ratio and the coherence time and can be calculated for AC magnetic fields via \cite{Taylor2008}:

Finally, the figure of merit of our scanning probes, the shot noise limited sensitivity to AC magnetic fields $\eta_{AC}$ can be calculated via:\cite{Taylor2008}
\begin{equation}
\eta_\mathrm{AC} \approx  \frac{\pi }{2  \gamma_{\rm NV} \mathrm{C} \sqrt{\mathrm{T_2}} }
  \label{eq:sensitivity}
   ,
\end{equation}
with $1/\mathrm{C}=\sqrt{1+2\left(\alpha_0+\alpha_1\right)/\left(\alpha_0-\alpha_1\right)^2}$. For the scanning probe measured in Fig.\,\ref{Fig:characterization} (e), we derive a sensitivity of $\mathrm{\eta_{AC}} \approx\,14\,\pm\,1 \,\mathrm{nT}/\sqrt{\mathrm{Hz} } $.
For $13$ scanning probes, we determined the magnetic field sensitivities as summarized in Fig. \ref{Fig:characterization}(f) and find an average sensitivity of  $\mathrm{\eta_{AC}} \approx\,50\,\pm\,20\,\,\mathrm{nT}/\sqrt{\mathrm{Hz}} $. The shot noise limited sensitivity to DC magnetic fields can be equivalently determined by using the relation $\mathrm{\eta_{DC}}=2/\pi \sqrt{\mathrm{T_2}/\mathrm{T_2^*}}\,\mathrm{\eta_{AC}}$.\cite{Taylor2008} Typical values for $T_2^*$ are few $\mu s$ and the resulting average DC sensitivity is therefore $\mathrm{\eta_{DC}}\approx\,200\,\mathrm{nT}/\sqrt{\mathrm{Hz}}$.\

\subsection{Transfer to scanning probe setup \label{sec:transfer}}
\begin{figure}
\includegraphics[width =8.6cm]{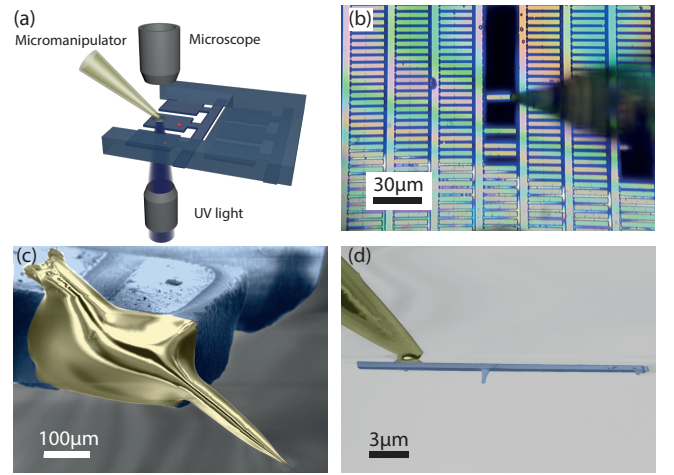}
\caption{(a) Schematic of the setup for gluing the scanning probe to quartz tips. (b) Optical microscope image during the transfer process. The scanning probe is glued to the apex of the quartz capillary tip using UV glue and the scanning probe is detached from the diamond chip by breaking. (c) SEM image of the scanning probe attached to one prong of a tuning fork. (d) SEM image of the final scanning probe attached to the end of the quartz tip.
\label{Fig:transfer}}
\end{figure}
In order to employ the scanning probes for imaging, the individually characterized cantilevers have to be transferred to an AFM head. Previous work employed ion beam assisted metal deposition to attach scanning probes to a quartz rod and subsequent focused ion beam (FIB) milling to detach the diamond scanning probe from the substrate.\cite{Maletinsky2012} This approach suffers from low yield, high complexity and significant contamination of the scanning probe by the Gallium ions used for FIB. Here we present an alternative method we developed to transfer the scanning probes using micromanipulators (Sutter Instruments, MPC-385) under ambient conditions. Using quartz micropipettes with an end diameter of $\sim3~\mu$m, we apply $\sim3~\mu$m sized droplets of UV curable glue (Thorlabs, NO81) to the device to be transferred [see Fig. \ref{Fig:transfer}\,(b)]. After curing the glue, we remove the device from the substrate by mechanically breaking the holding bar [0.5 $\mu$m wide, see e.g. Fig. \ref{Fig:characterization}\,(a)] with the quartz pipette.

In a second step, we glue the quartz tip with the scanning probe to a tuning fork attached to an AFM head [see Fig. \ref{Fig:transfer} (c)]. To that end, we employ a stereo microscope setup which allows precise alignment of the scanning probe with respect to the AFM head and subsequent gluing of the quartz tip to the tuning fork using UV curable optical glue. As a last step, we carefully break the the quartz pipette above its connection (gluing point) to the tuning fork using a diamond scribe [see Fig. \ref{Fig:transfer} (c)].

With this procedure, we are able to produce tuning fork based AFM heads with the scanning probes aligned within a few degrees to the AFM holder in a robust and fast way. The UV glue forms a strong connecting link that can be used even in cryogenic environment\cite{Thiel2015} and enables long-term use of the device.

\section{Nanoscale scanning probe magnetometry\label{sec:mag}}
\begin{figure}
\includegraphics[width =8.6cm]{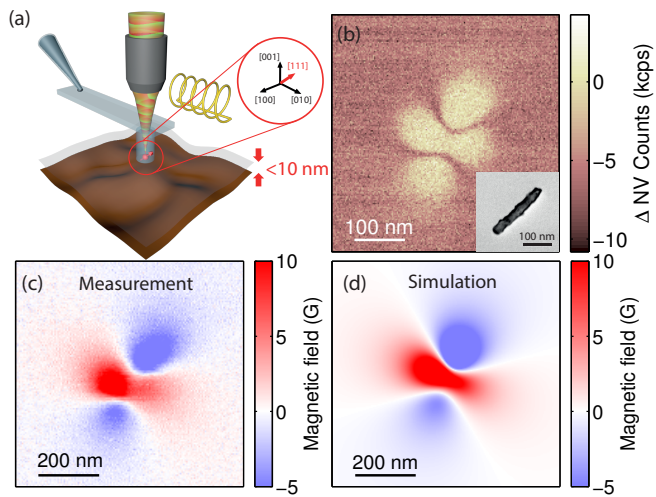}
\caption{(a) Schematic of the combined confocal and AFM setup. The sample is scanned in close proximity under the NV center and the magnetic field dependent fluorescence is collected using a confocal microscope. (b) Iso-magnetic-field image of a single Ni nanorod; negative fluorescence contrast indicates a local field smaller than the detection threshold (see text); the inset shows a SEM image of a rod. (c) Full field map of a single Ni nanorod and (d) the magnetic field of a point dipole ($m=3.75\times10^{-17}$A/m$^2$) projected onto the NV axis for an NV center located $80\,$nm above the dipole.
\label{Fig:magnet}}
\end{figure}

 We now demonstrate the performance of scanning quantum sensor by showing our device's capability for quantitatively imaging  magnetic fields with nanoscale resolution. Our setup, consisting of a combined AFM and confocal microscope, has been described elsewhere.\cite{Appel2015}

We applied NV magnetometry to study single Ni nanorods. These nanorods have various potential applications such as magneto-optical switches \cite{Klein2009} or as probe particles in homogeneous immunoassays for the detection of proteins\cite{Schrittwieser2013} and in microrheology.\cite{Tschoepe2014} NV center based magnetometry allows us to study the magnetic properties (spin densities, spin textures etc.\cite{Hingant2015a,Tetienne2015a}) of individual particles. Here, we present two different approaches for imaging the stray field of single Ni nanorods which have typical diameters $\sim 24\,\mathrm{nm}$ and lengths $\sim 230\,\mathrm{nm}$ and which are deposited from a solution onto a quartz substrate [see inset of Fig. \ref{Fig:magnet}(b)].\

Our first imaging method is based on measuring isomagnetic field lines.\cite{Maletinsky2012} For this purpose, we fix the MW frequency to the NV spin transition frequency as determined in the absence of the sample. In the presence of a magnetic field, e.g. the stray field of the Ni nanorod, the frequency of NV spin transition gets detuned from the MW frequency which results in an increase of NV fluorescence [see Fig.\, \ref{Fig:characterization}(b)]. While scanning the NV spin at a distance $d$ over the sample, the iso-field line at zero magnetic field is therefore mapped onto decreased NV fluorescence [Fig. \ref{Fig:magnet}\,(b)]. Such isomagnetic field imaging is a fast method for probing nanomagnetic structures and their dynamics.\cite{Tetienne2014} \

 For a complete analysis of the magnetic stray field of the nanorod, it is necessary to perform full, quantitative magnetic stray field mapping. To that end, the Zeeman shift induced by the magnetic field needs to be detected. Various methods to measure the Zeeman shift have been discussed.\cite{Tetienne2015a,Schoenfeld2011,Haeberle2013} We pursue the approach presented in Ref.\,\onlinecite{Schoenfeld2011}. A feedback loop is used to lock the MW frequency to the NV spin transition frequency. Using such a frequency lock, the magnetic field can be measured while scanning the NV sensor over the sample. Figure\,\ref{Fig:magnet} (c) depicts the full stray field of the Ni nanorod obtained via such a frequency feedback loop.

 The measured stray field matches the stray field expected for a single dipole. Assuming a point dipole with a magnetic moment of $m=3.75\times10^{-17}$A/m$^2$, as measured for similar rods with different methods,\cite{Schrittwieser2013} we calculated the magnetic field projected onto the NV axis. With this method we find agreement between measurement and model and estimate a distance of $\sim70\,$nm  between the sample surface and NV center. This distance sets the spatial resolution of the presented scanning magnetometer. The NV center can in principle detect changes of magnetic fields on length scales of $\sim1\,$nm, set by the spatial extent of its electronic wavefunction. Consequently, the imaging resolution of our NV magnetometer is not limited by the detector size but solely by the NV-to-sample distance. We emphazise that the distance of $\sim70\,$nm we determined is a rough estimate and a more precise model has to be used to explain in detail the magnetic field profile. Factors that contribute to this larger-than-expected distance include a polymer layer of unknown thickness surrounding the nanorods,\cite{Guenther2011} a potential water-layer that typically covers samples under ambient conditions, or dirt sticking to the tip and acting as an additional spacing layer. In the absence of such factors we observed NV-to-sample distances between $10$ and $25\,$nm (see  Ref.\,\onlinecite{Appel2015} and  Ref.\,\onlinecite{Thiel2015}) certifying the nanoscale resolution our scanning probes offers.

\section{Discussion and Perspectives}
The all-diamond scanning probes we fabricated have proven their potential for detecting magnetic fields with high sensitivity and nanoscale resolution. We conclude by highlighting improvements, which are currently investigated to increase the performance of the presented scanning probe technique.

 To increase the sensitivity of the scanning probes with single NV centers, a long coherence time $T_2$ and high fluorescence rates are required as can be seen in Eq.\,\ref{eq:sensitivity}. Thus, efficiently collecting the NV's fluorescence is crucial for highly sensitive scanning probes. Using our $200\,$nm diameter, cylindrical pillar, we increase the typical fluorescence count rates by a factor of $\sim 3$ compared to bulk diamond. More complex photonic geometries such as tapered pillars \cite{Momenzadeh2015} are currently investigated to further enhance the collection efficiency and might be useful for scanning probes. Further improvements are also expected by optimizing the crystal orientation of the employed diamond samples. Here we employ (100) oriented diamond which is the standard orientation of commercially available high purity diamond. However, in (111) oriented diamond, the NV axis can be oriented perpendicularly to the diamond surface, which yields improved photonic properties as compared to (100) oriented nanopillars.\cite{Neu2014}

A central advantage of our scanning probes is the use of high purity diamond which in principle allows long T$_2$ times to be reached. Unfortunately high resolution imaging requires NV centers in close proximity to the surface, which typically comes at the expense of shorter coherence times due to proximal surface spins.\cite{Romach2015} For the presented scanning probes, we have chosen an implantation depth of $9\pm4$\,nm which yields coherence times of $T_2=76\pm19\,\mu$s in the diamond plate before nanofabrication of the scanning probes. In our scanning nanopillars however, we find an average $T_2=44\pm26\,\mu$s. We ascribe this reduction of coherence to unwanted and currently unknown surface defects which are created on the diamond surface during etching. Recent work\cite{Oliveira2015} suggests that a low bias, 'soft' oxygen plasma can remove such plasma induced surface damage and could thereby provide a remedy for this problem. This and similar methods\cite{Cui2013,Osterkamp2013,Lovchinsky2016} still remain to be tested on diamond scanning probes and their influence on NV spin coherence times remains an open question.

Another challenge is the creation of NV centers with a controlled distance to the diamond surface in the nanometer range. The ion implantation employed here partly suffers from a low yield ($<$ 1\%) and a significant uncertainty in the resulting NV depth ($9\pm4$ nm). Recent work suggests '$\delta$-doping' as an alternative: in this technique down to 2 nm thin, nitrogen enriched layers are engineered during the growth of diamond\cite{Ohno2012,Ohno2014}. However, creating the necessary density of NV centers sufficient to yield one NV per pillar still remains an outstanding challenge.\cite{Ohno2014}.

The presented fabrication process is suited for structuring arrays with hundreds of scanning probes. We so far used 50 $\mu$m thin diamond plates and handled them without any permanent bonding to a carrier system. However, permanent bonding to Si carriers as e.g. described in \cite{RiedrichMoeller2015,Tao2013} using HSQ e-beam resist might potentially enable the use of thinner diamond plates and structuring of even more device arrays in a single step. Bonding to carriers might potentially facilitate sample handling, enhance the device yield
and pave the way towards further scaling of the presented fabrication processes.

\section{Conclusion}
In this paper, we described in detail our advanced fabrication process for all-diamond scanning probes starting from commercially available diamond material. We demonstrated the efficient integration of our tips into a generic AFM setup and imaged the dipolar magnetic field of Ni nanorods with two different measurement techniques. Our state of the art scanning probes, with the NV-center placed $\sim 10\,$nm below surface of the scanning pillar, have sensitivities of $\mathrm{\eta_{AC}}\approx50\pm20\,\mathrm{nT}/\sqrt{\mathrm{Hz}} $. Finally, we highlight future avenues to push NV center based magnetometry to its ultimate limit to yield scanning NV magnetometers capable of detecting weak magnetic signal down to small ensembles of nuclear spins.\cite{Mamin2013,Ajoy2015} \\

\section*{Acknowledgments}
We thank B. Shields and D. Rohner for fruitful discussions, J. Teissier for assistance with nanofabrication, L. Thiel for support with the experiment control software and A. Kretschmer for creating the illustrations. We gratefully acknowledge financial support through the NCCR QSIT, a competence center funded by the Swiss NSF, and through SNF Grant No. 142697 and 155845. This research has been partially funded by the European Commission's 7.\ Framework Program (FP7/2007-2013) under grant agreement number 611143 (DIADEMS). EN acknowledges funding via the NanoMatFutur program of the german ministry of education and research.

% Create the reference section using BibTeX: Bibliography appended in the end
\bibliographystyle{apsrev4-1}

\end{document}